\documentclass{webofc}
\usepackage[varg]{txfonts}   
\usepackage{subcaption}
\usepackage{float}

\newcommand{\geant}{\texttt{Geant4} }
\begin{document}
\title{Generative Adversarial Networks for LHCb Fast Simulation}
%
%

\author{\firstname{Fedor} \lastname{Ratnikov}\inst{1}\fnsep\thanks{\email{fedor.ratnikov@cern.ch}} 
  }

\institute{National Research University Higher School of Economics, Moscow, Russia} 

\abstract{%
  LHCb is one of the major experiments operating at the Large Hadron Collider at CERN. The richness of the physics program and the increasing precision of the measurements in LHCb lead to the need of ever larger simulated samples. This need will increase further when the upgraded LHCb detector will start collecting data in the LHC Run 3. Given the computing resources pledged for the production of Monte Carlo simulated events in the next years, the use of fast simulation techniques will be mandatory to cope with the expected dataset size. Generative models, which are nowadays widely used for computer vision and image processing, are being investigated in LHCb to accelerate generation of showers in the calorimeter and high-level responses of Cherenkov detector. We demonstrate that this approach provides high-fidelity results and discuss possible implications of these results. We also present an implementation of this algorithm into LHCb simulation software and validation tests.
}
\maketitle

\section{Introduction}
\label{intro}
Detailed simulation of the detector response on different types of
physics events is a vital component of every experiment in high energy
physics. Without such simulation it is virtually impossible to infer
a physics result from the experimental observations. Detailed
simulation however requires significant computing resources. Moreover,
 simulation is the primary consumer of computing resources:
about 80\% of the total computing  is used by  HEP experiment for simulation.

\begin{figure}[htb!]
\centering
\includegraphics[width=0.95\textwidth]{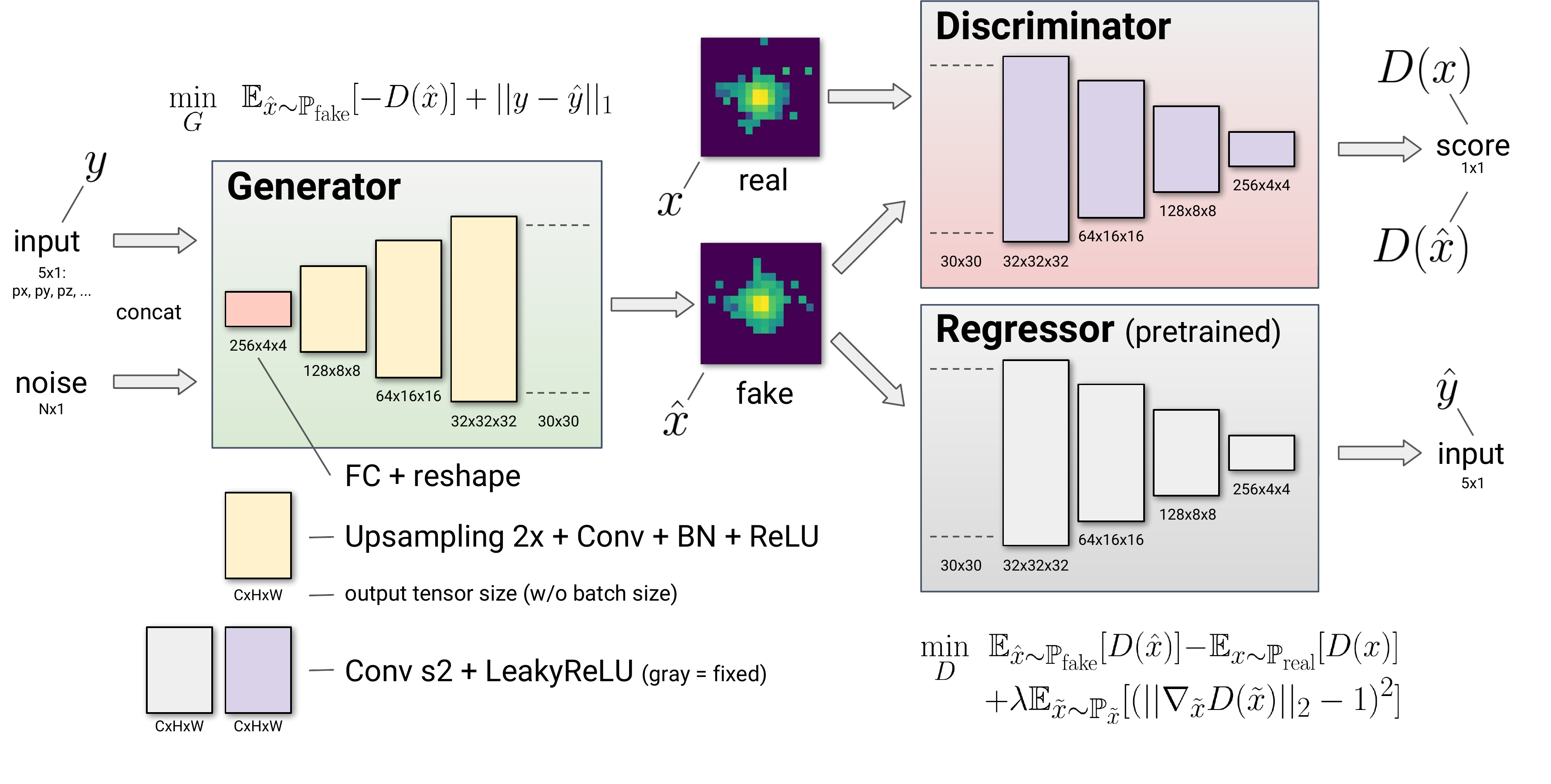}
\caption{Model architecture. Pre-trained regressor for the particle parameters prediction makes our model conditional. Thanks to building up the information from the pre-trained regressor into the discriminator gradient we learn $G$ to produce a specific calorimeter response.}\label{fig:model}
\end{figure}

\begin{figure}[htb!]
\captionsetup[subfigure]{justification=centering}
  \centering
  \begin{subfigure}{0.24\textwidth}
    \centering
    \includegraphics[width=1\textwidth]{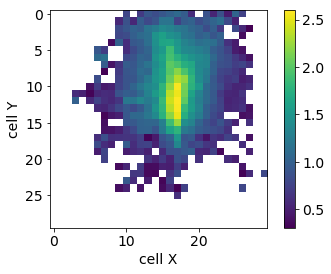}
  \end{subfigure}
  \begin{subfigure}{0.24\textwidth}
    \centering
    \includegraphics[width=1\textwidth]{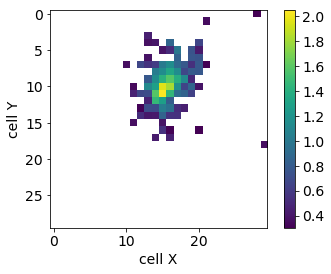}
  \end{subfigure}
    \begin{subfigure}{0.24\textwidth}
    \centering
    \includegraphics[width=1\textwidth]{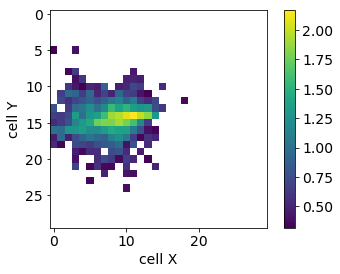}
  \end{subfigure}
  \begin{subfigure}{0.24\textwidth}
    \centering
    \includegraphics[width=1\textwidth]{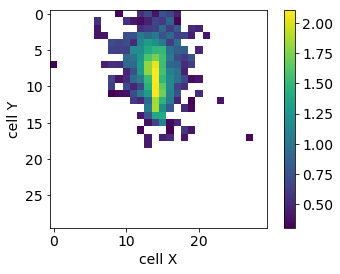}
  \end{subfigure}\\
   \begin{subfigure}{0.24\textwidth}
    \centering
    \includegraphics[width=1\textwidth]{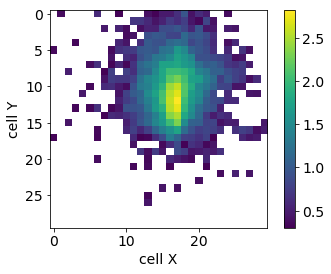}
    \caption{\\$E_0 = 63.7~\text{GeV}$ }
  \end{subfigure}
  \begin{subfigure}{0.24\textwidth}
    \centering
    \includegraphics[width=1\textwidth]{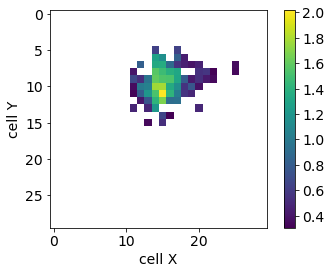}
    \caption{\\$E_0 = 6.5~\text{GeV}$ }
  \end{subfigure}
    \begin{subfigure}{0.24\textwidth}
    \centering
    \includegraphics[width=1\textwidth]{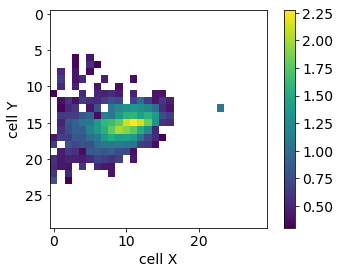}
    \caption{\\$E_0 = 15.6~\text{GeV}$ }
  \end{subfigure}
  \begin{subfigure}{0.24\textwidth}
    \centering
    \includegraphics[width=1\textwidth]{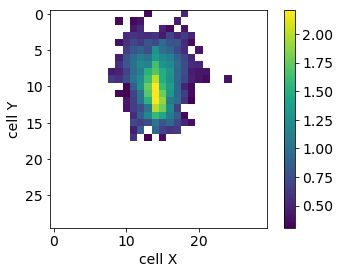}
    \caption{\\$E_0 = 15.9~\text{GeV}$ }
  \end{subfigure}
 
  \caption{Visual comparison of generated calorimeter showers. Showers generated with \geant (first row) and the showers,
    simulated with our model (second row) for four different sets of
    input parameters. Colour represents $log_{10}(\frac{E}{MeV})$ for every cell.}
  \label{fig:geant_vs_ours}
\end{figure}

\begin{figure}[htb!]
  \centering
  \begin{subfigure}[t]{0.3\textwidth}
    \centering
    \includegraphics[width=1\textwidth]{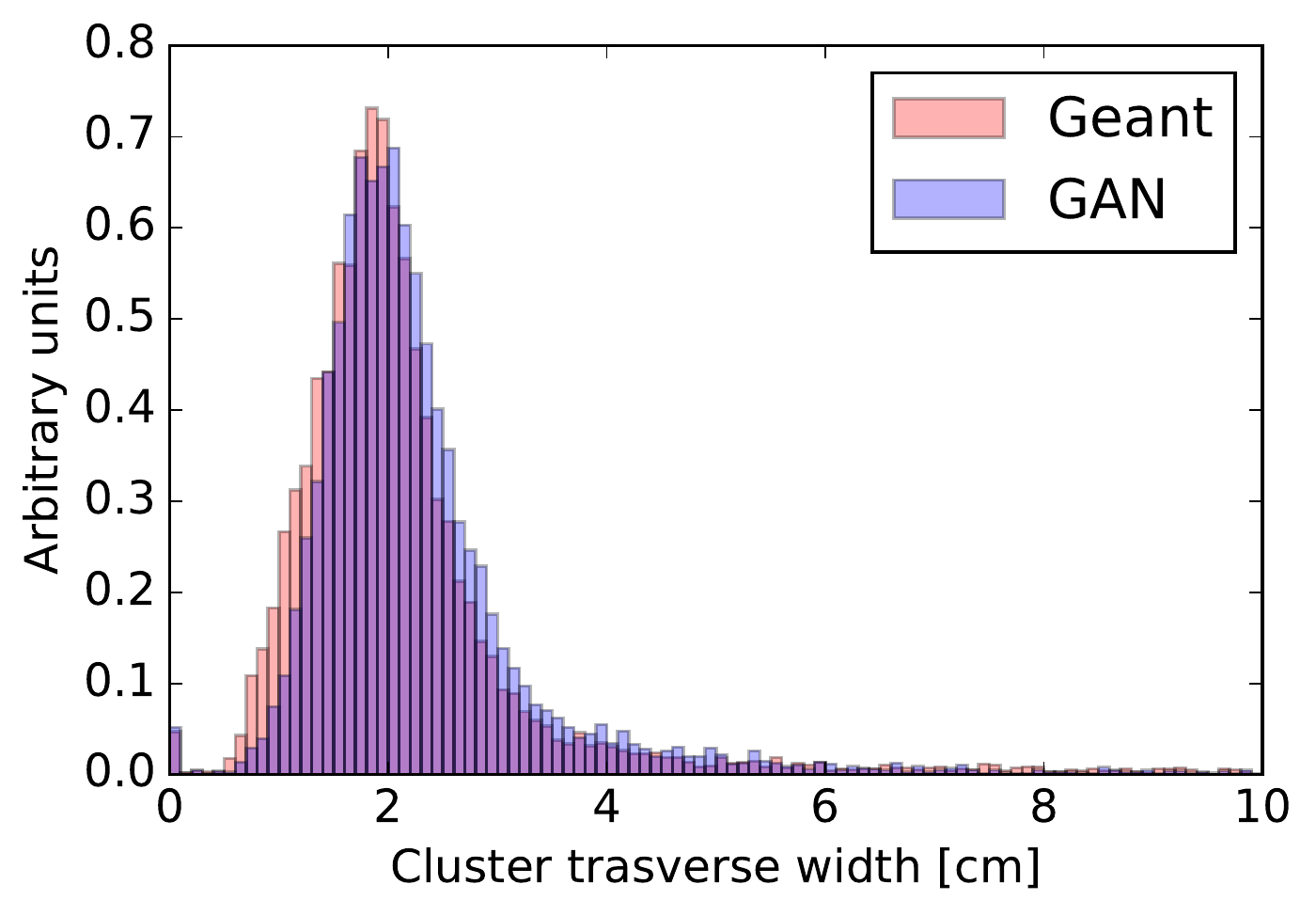}
    \caption{The transverse width of clusters.}
  \end{subfigure}\hspace{0.2\textwidth}
 \begin{subfigure}[t]{0.3\textwidth}
    \centering
    \includegraphics[width=1\textwidth]{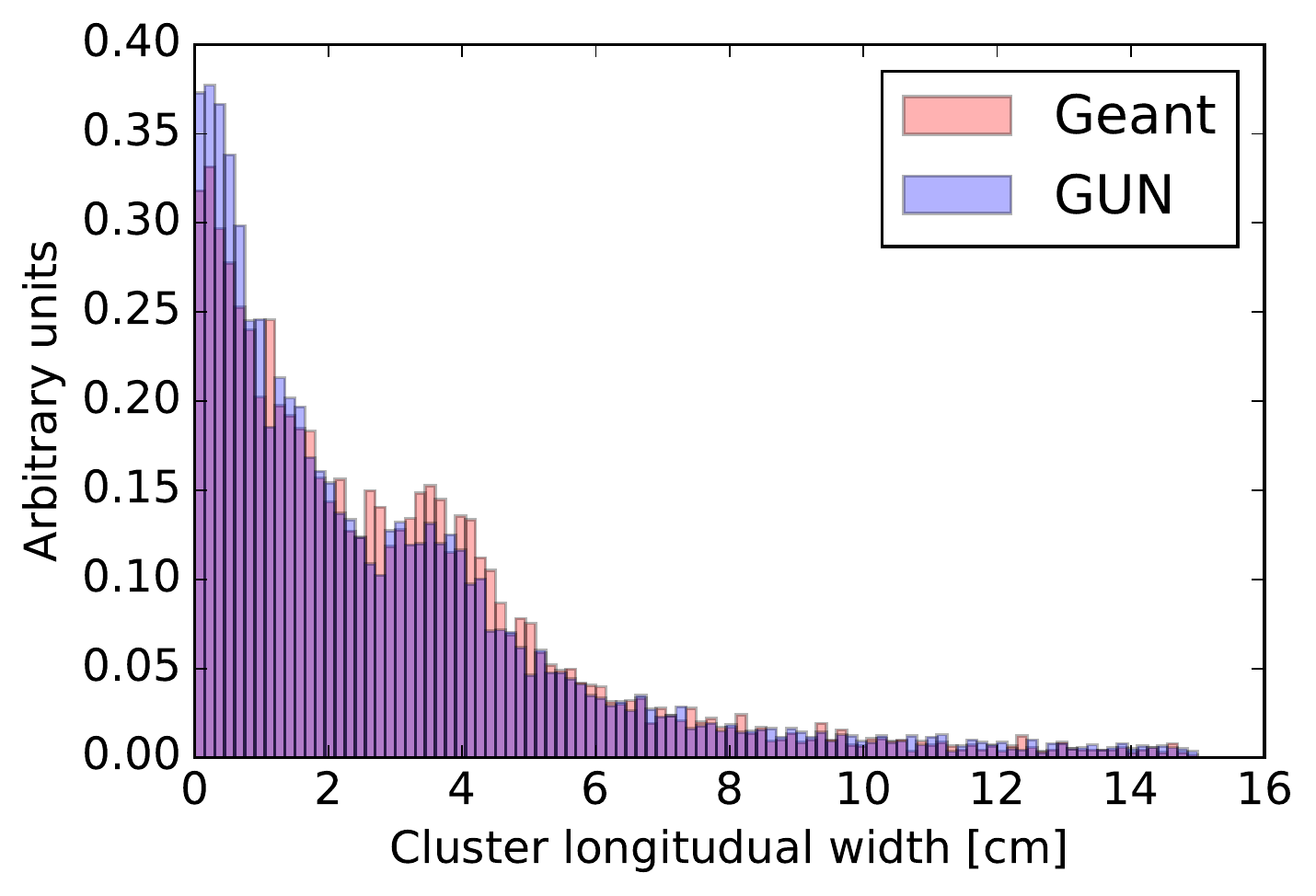}
    \caption{The longitudinal width of clusters.}
  \end{subfigure}
  \begin{subfigure}[t]{0.3\textwidth}
    \centering
    \includegraphics[width=1\textwidth]{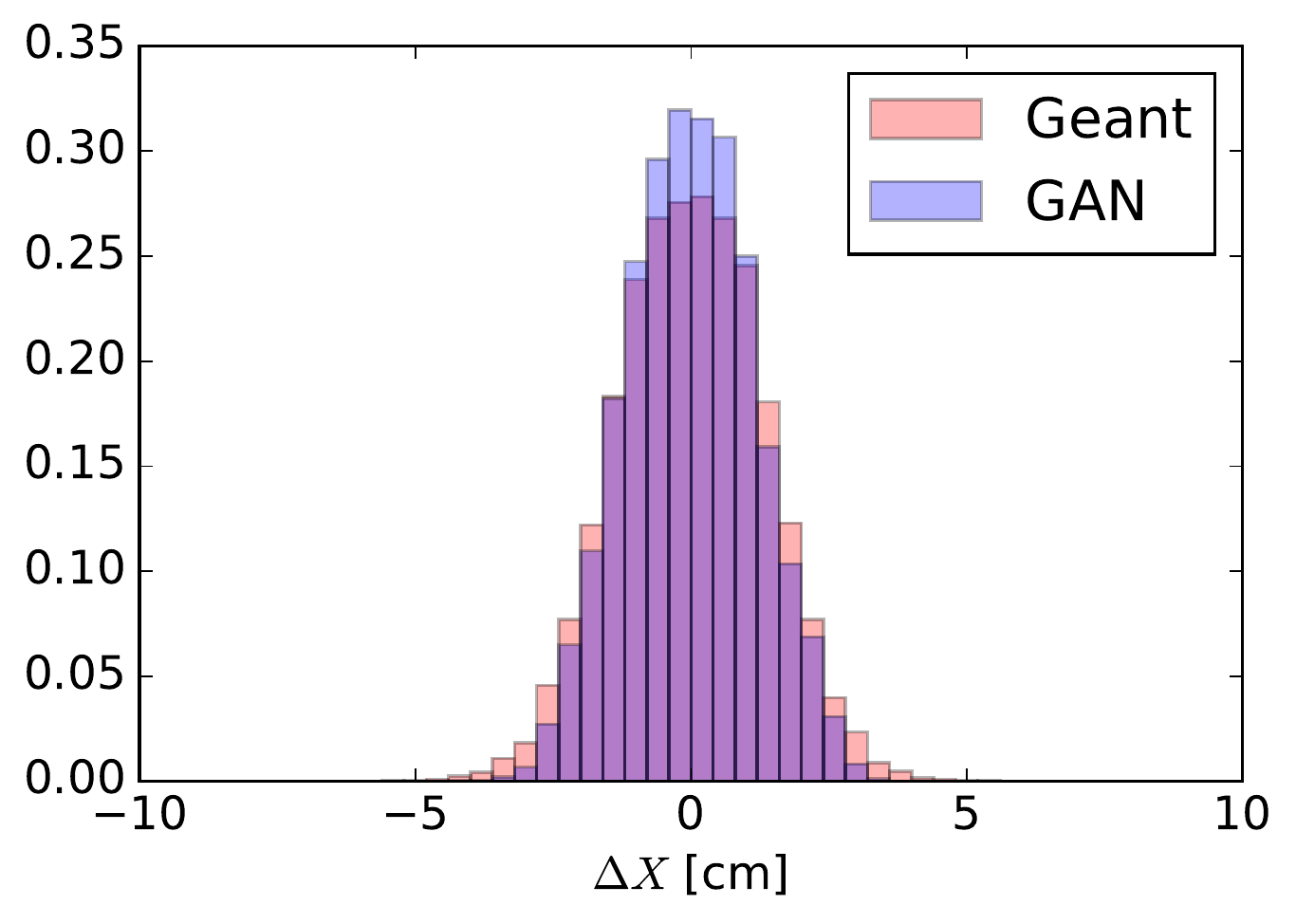}
    \caption{$\Delta X$ between cluster centre of mass and the true particle coordinate.}
  \end{subfigure}\hspace{0.2\textwidth}
  \begin{subfigure}[t]{0.3\textwidth}
    \centering
    \includegraphics[width=1\textwidth]{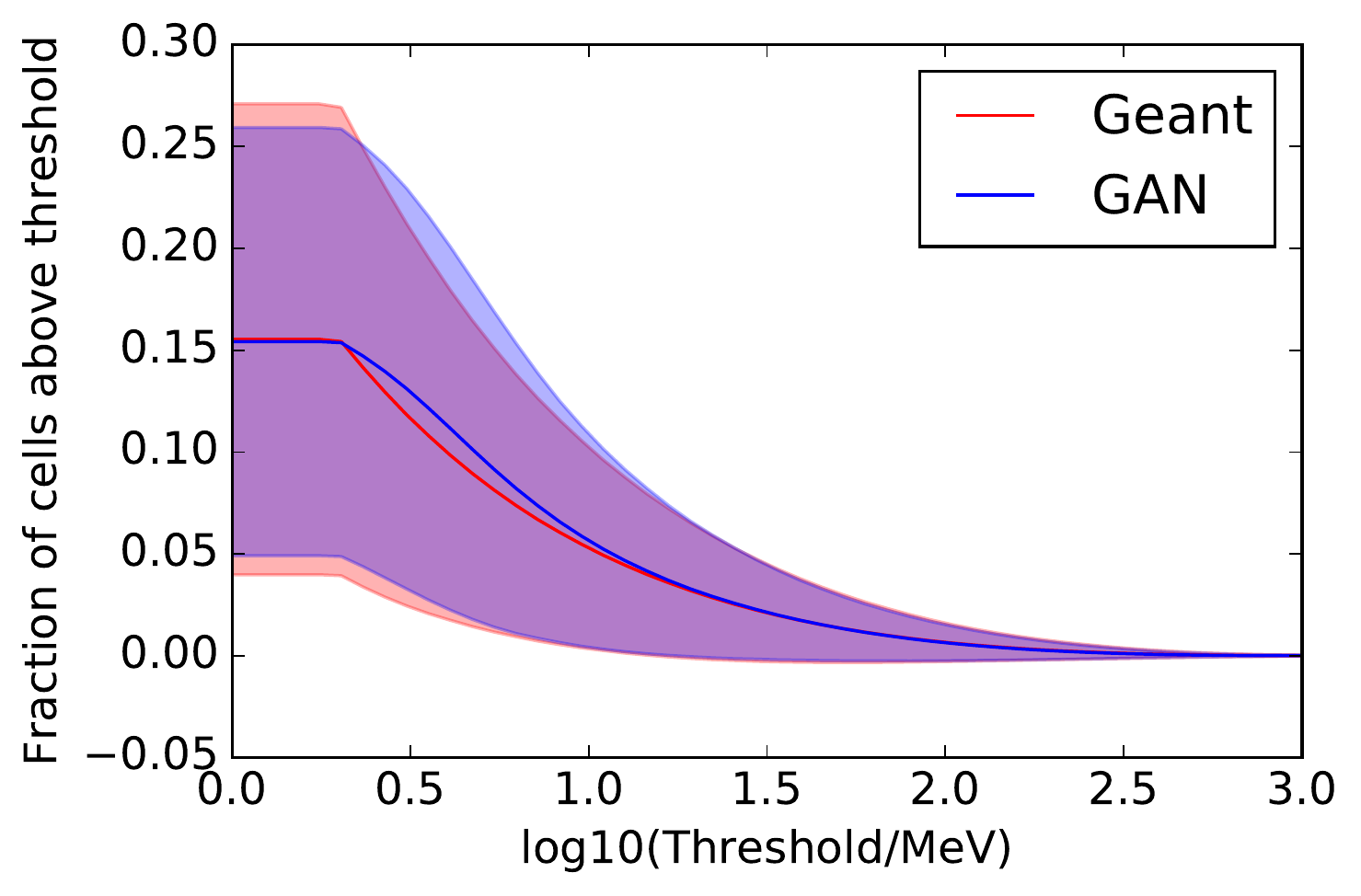}
    \caption{The sparsity in the 30x30 cells matrix containing clusters.}
  \end{subfigure}
  \begin{subfigure}[t]{0.3\textwidth}
    \centering
    \includegraphics[width=1\textwidth]{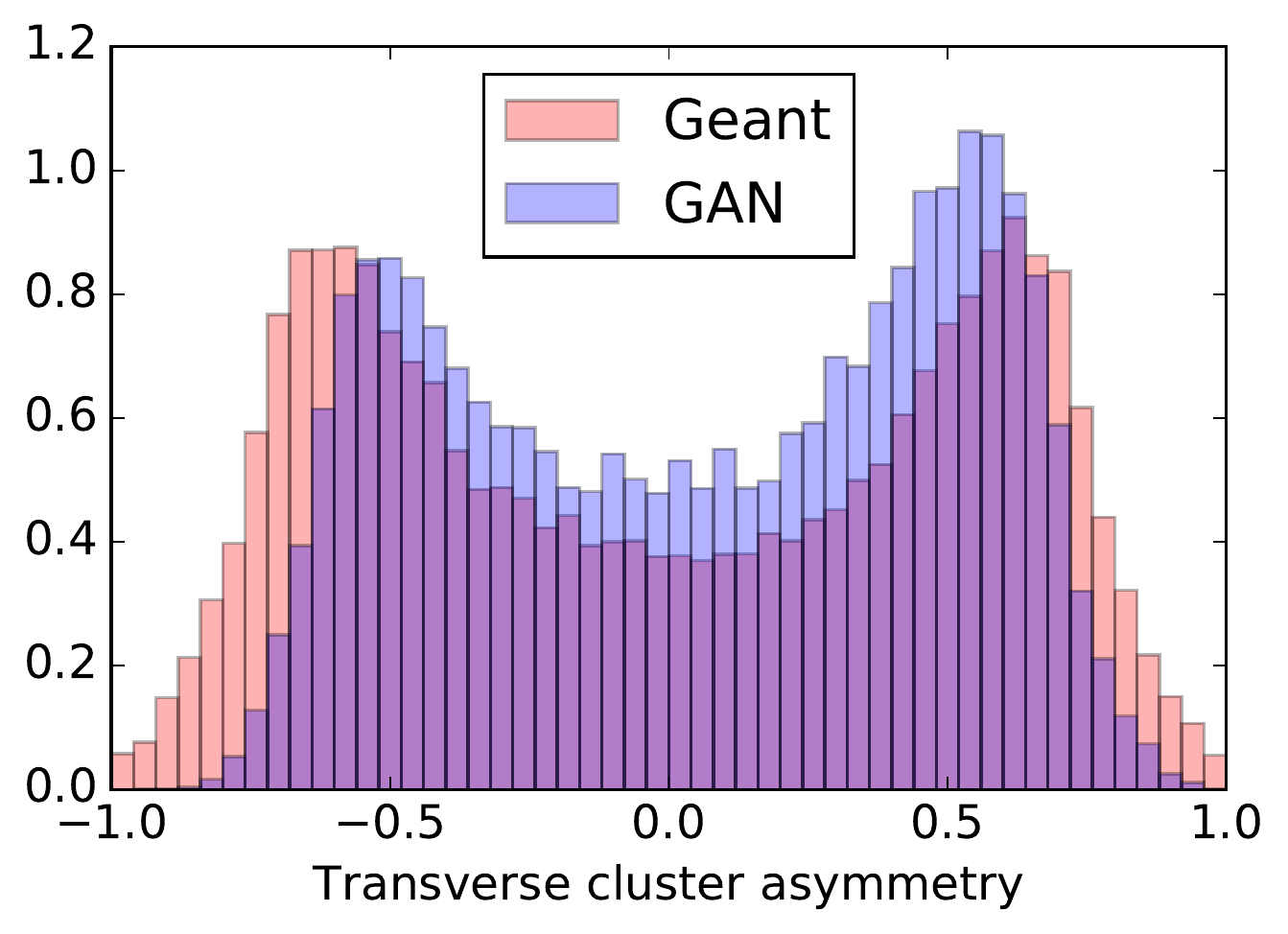}
    \caption{The transverse asymmetry of clusters.}
  \end{subfigure}\hspace{0.2\textwidth}
  \begin{subfigure}[t]{0.3\textwidth}
    \centering
    \includegraphics[width=1\textwidth]{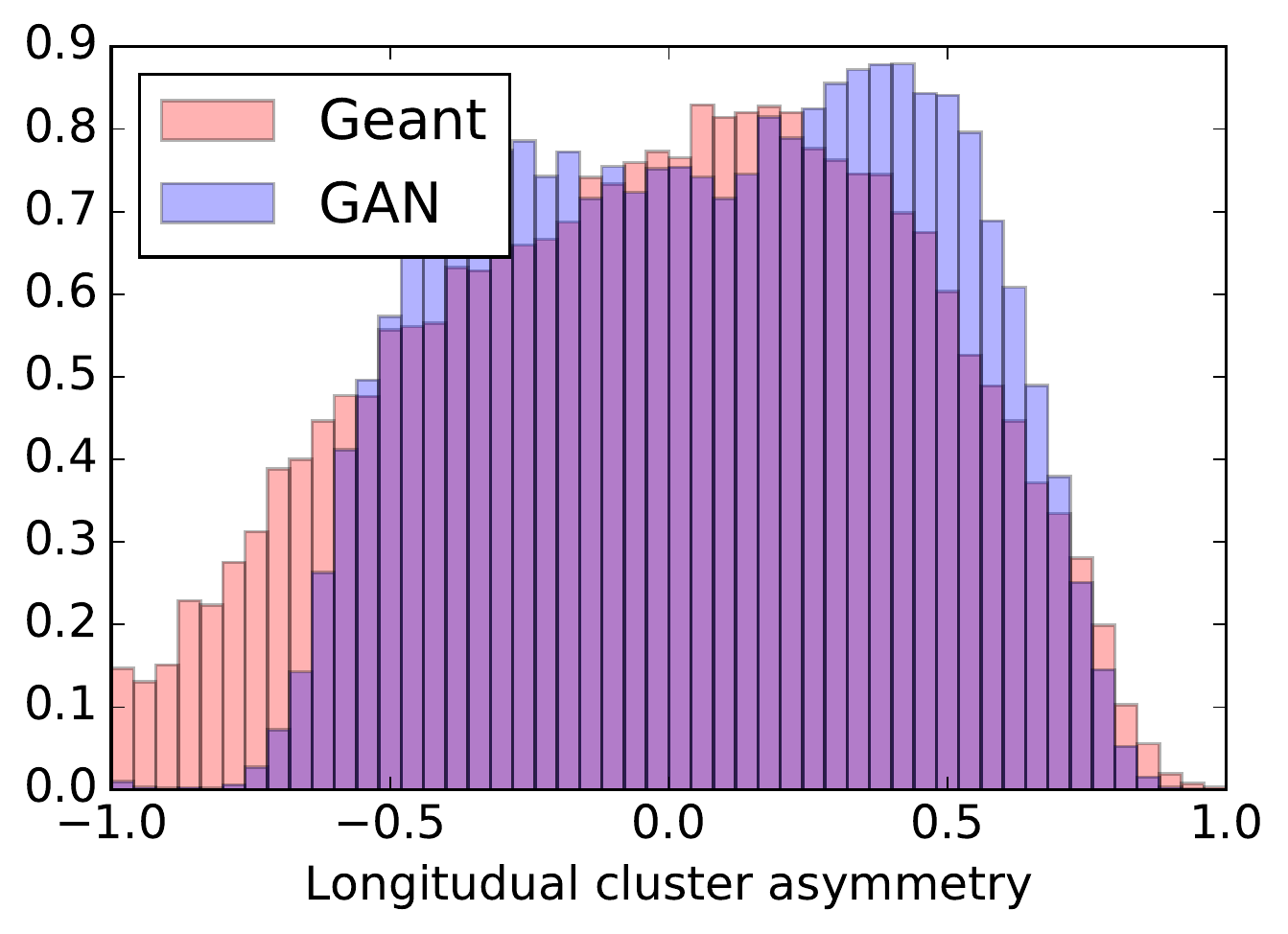}
    \caption{The longitudinal asymmetry of clusters.}
  \end{subfigure}
  \caption{Comparing physics characteristics for \geant simulated (red) and generated (blue) clusters.
   }
  \label{fig:ganquality}  
\end{figure}

\begin{figure}[htb!]
\centering
\includegraphics[width=0.6\textwidth]{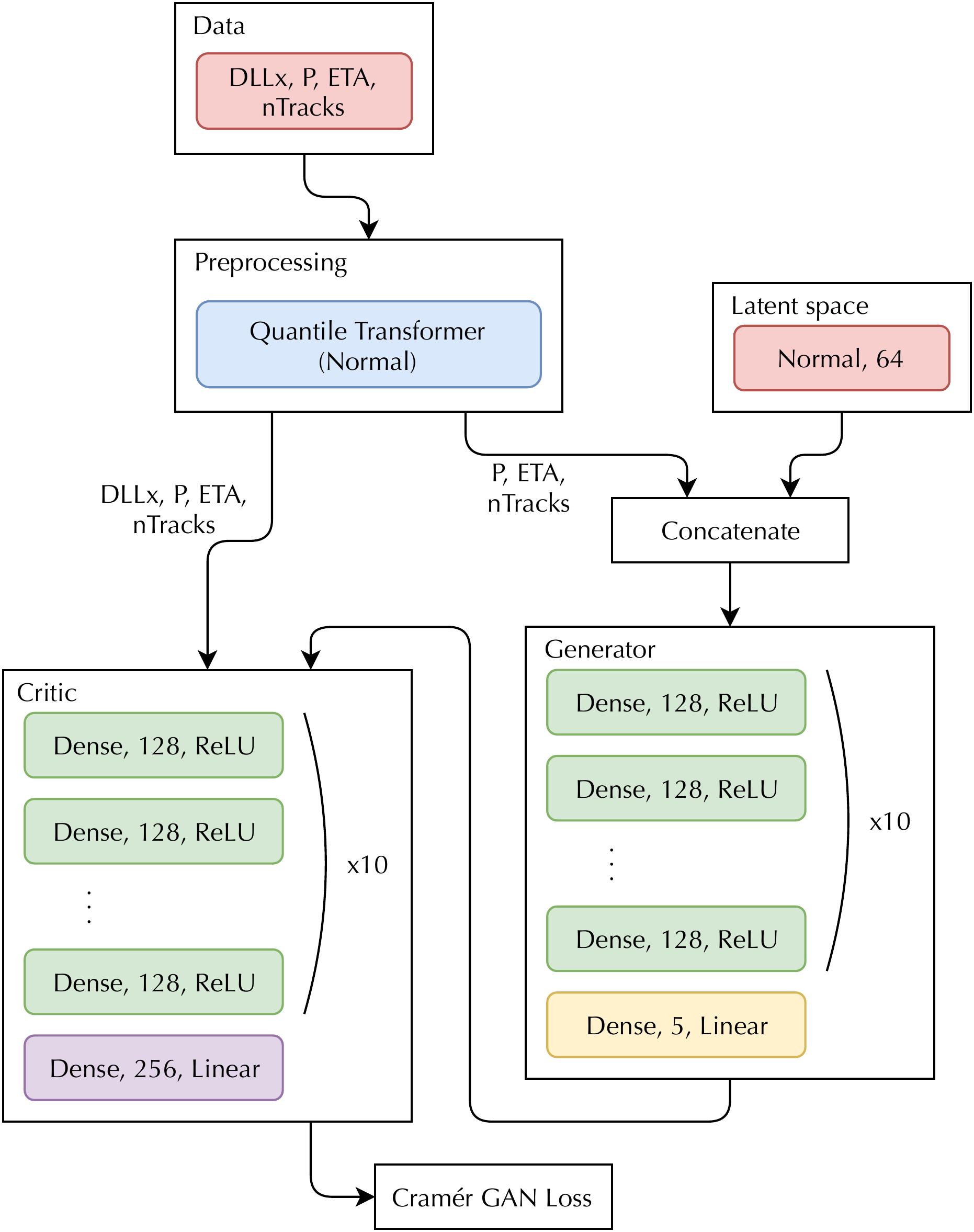}
\caption{RICH GAN model architecture. Generator is trained to produce five particle identification responses for the RICH detector for three input parameters: particle momentum, pseudorapidity, and the total event track multiplicity.}
\label{fig:richgan}       
\end{figure}

\begin{figure}[htb!]
    \centering
    \includegraphics[width=0.7\textwidth]{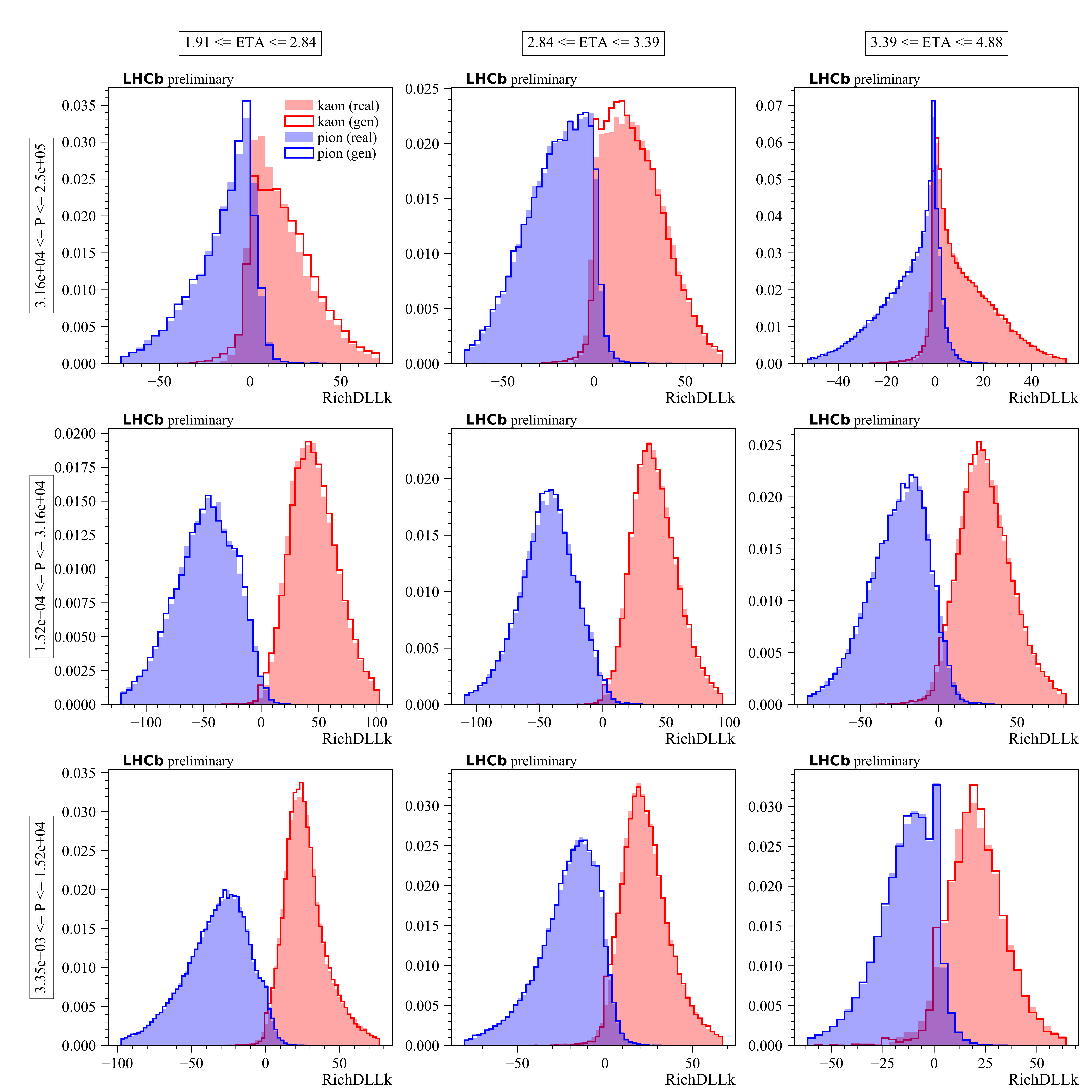}
    \caption{Visual comparison of RICH GAN quality. Weighted real data (filled histograms) and generated (open histograms) distributions of \texttt{RichDLLk} for kaon (red) and pion (blue) track candidates in bins of pseudorapidity (ETA) and momentum (P) over full phase-space. }
    \label{fig:RichDLLk}
\end{figure}

\begin{figure}[htb!]
    \centering
   \begin{subfigure}[t]{0.3\textwidth}
    \centering
    \includegraphics[width=1\textwidth]{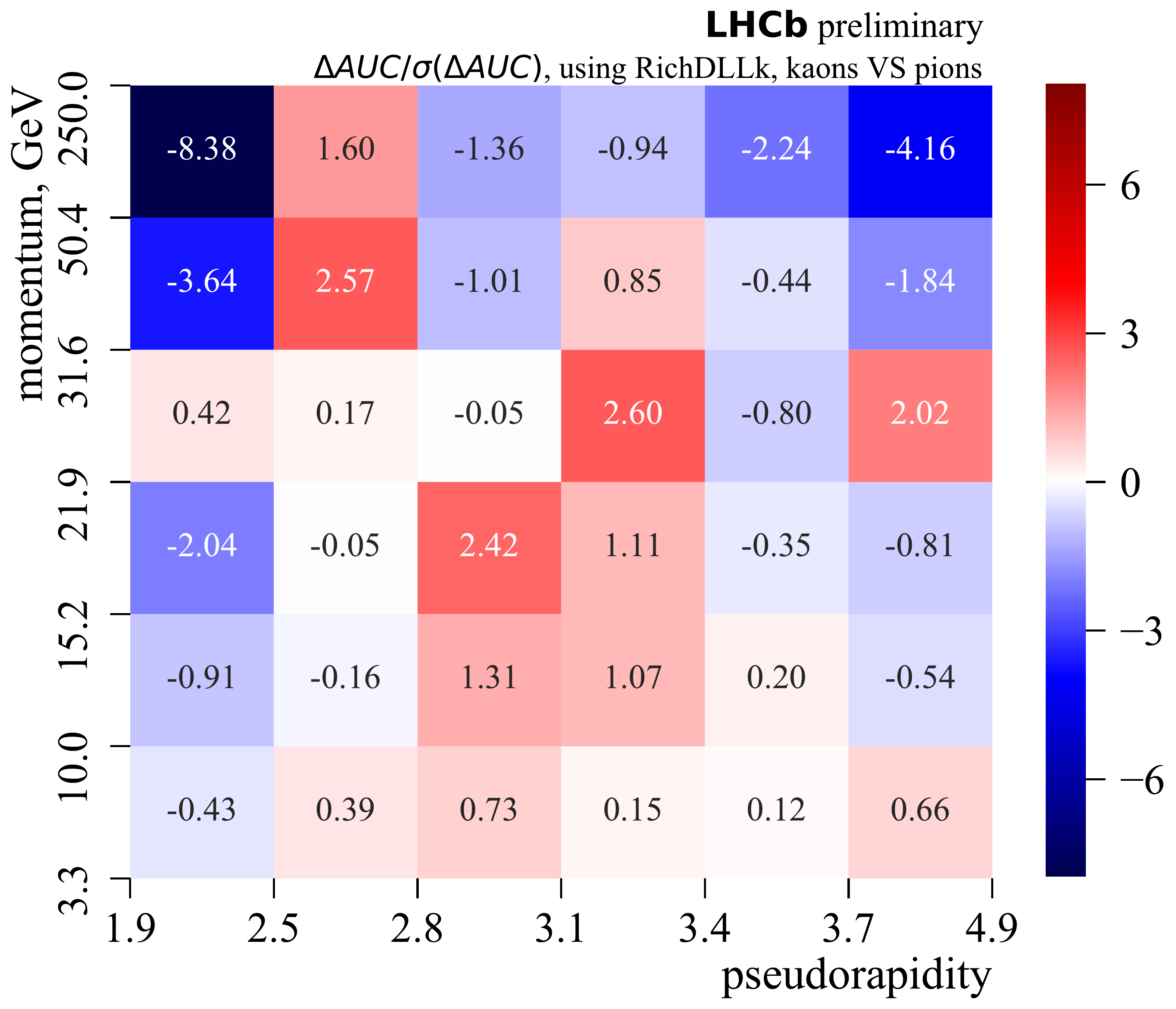} 
    \caption{kaons vs pions  }
  \end{subfigure}
   \begin{subfigure}[t]{0.3\textwidth}
    \centering
    \includegraphics[width=1\textwidth]{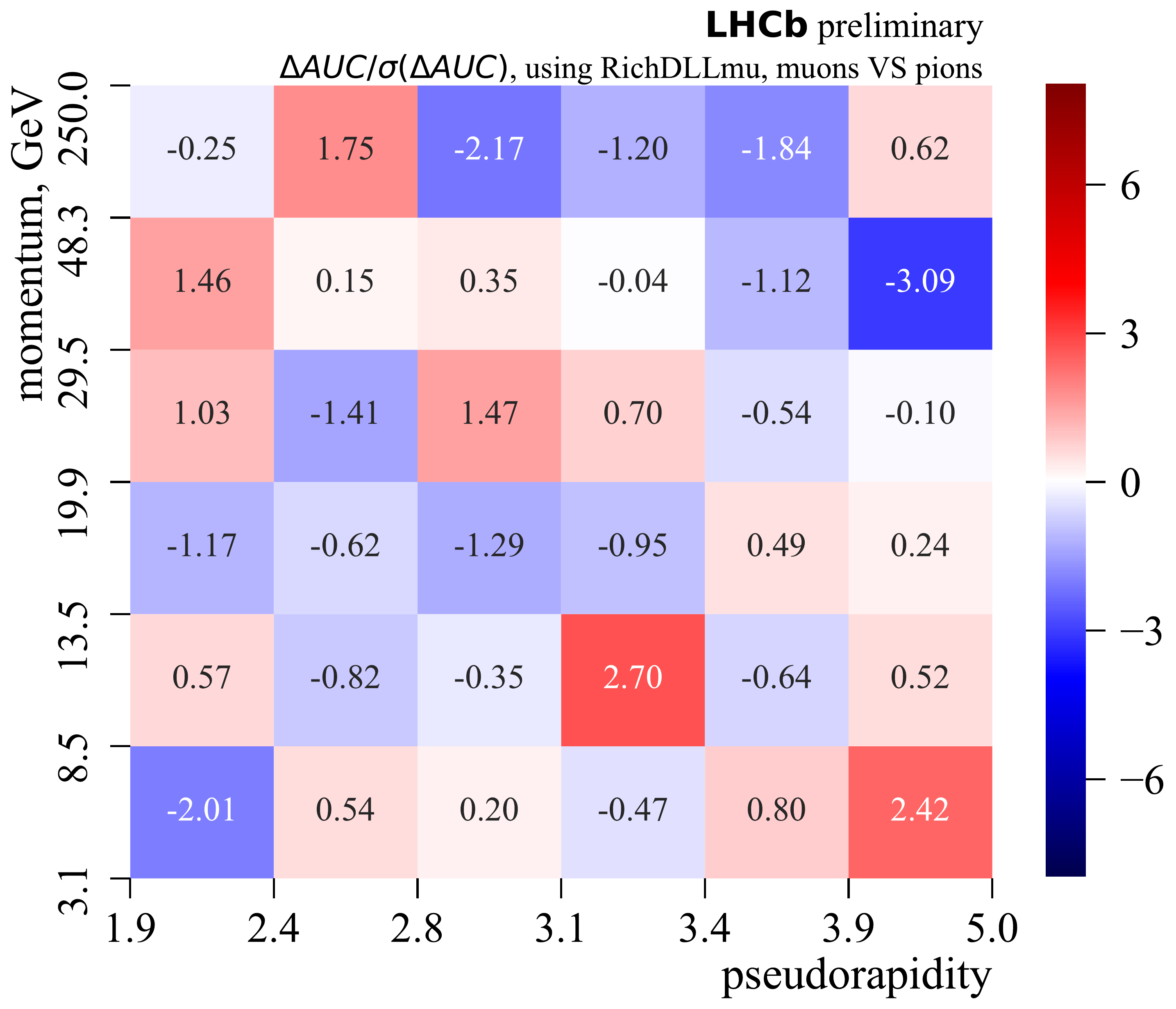}
    \caption{muons vs pions  }
  \end{subfigure}
  \begin{subfigure}[t]{0.3\textwidth}
    \centering
    \includegraphics[width=1\textwidth]{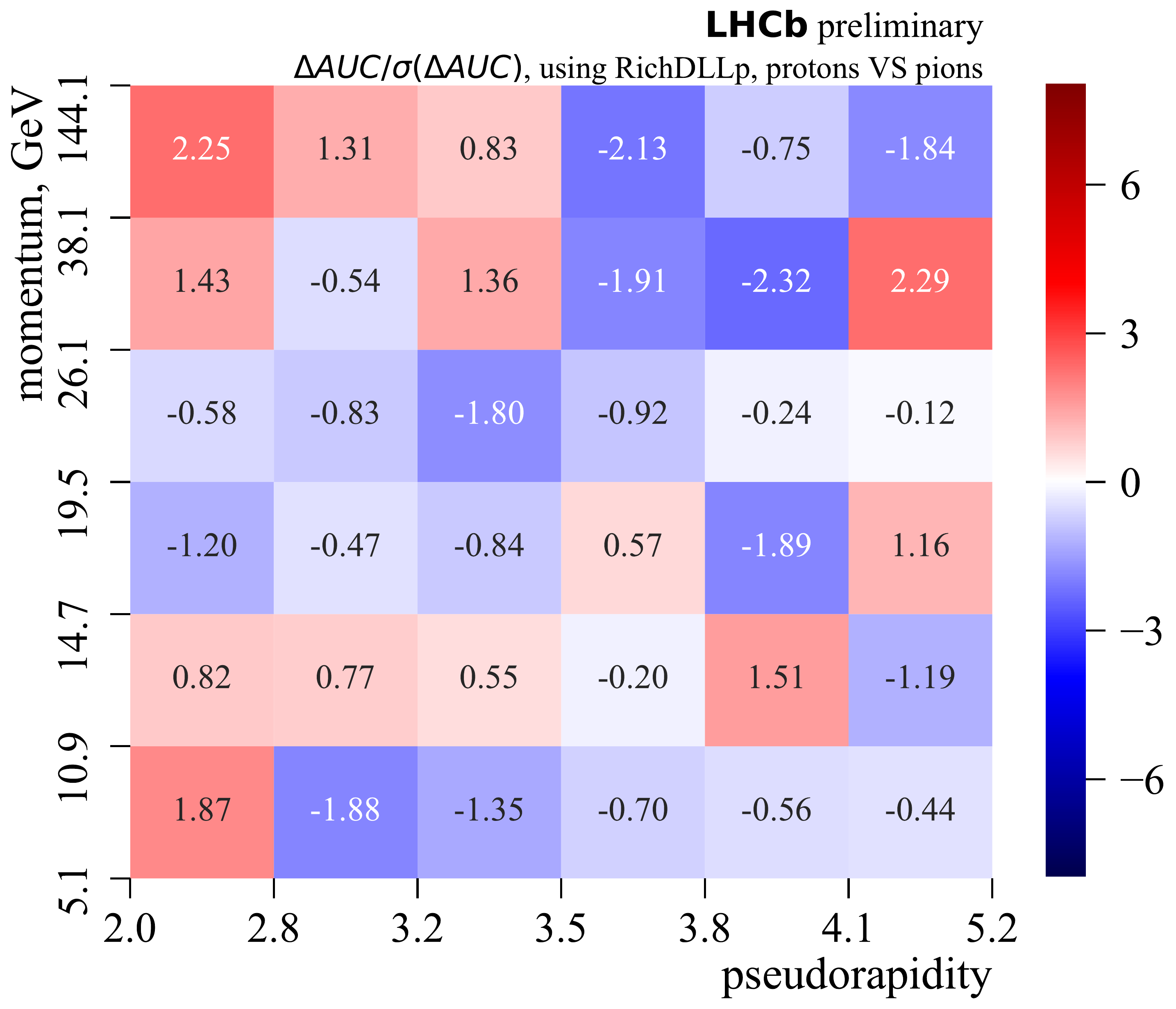} 
    \caption{protons vs pions  }
 \end{subfigure}
  \caption{Separation power comparison. Differences between ROC AUC for discriminating pions from kaons, muons and protons classifiedd with the \texttt{RichDLLk}, \texttt{RichDLLmu} and \texttt{RichDLLp} variables, respectively for real calibration samples and using RICH GAN generated values, divided by the ROC AUC uncertainties, in bins of momentum and pseudorapidity.}
    \label{fig:DAUCE}
\end{figure}

\begin{figure}[htb!]
    \centering
   \begin{subfigure}[t]{0.3\textwidth}
    \centering
    \includegraphics[width=1\textwidth]{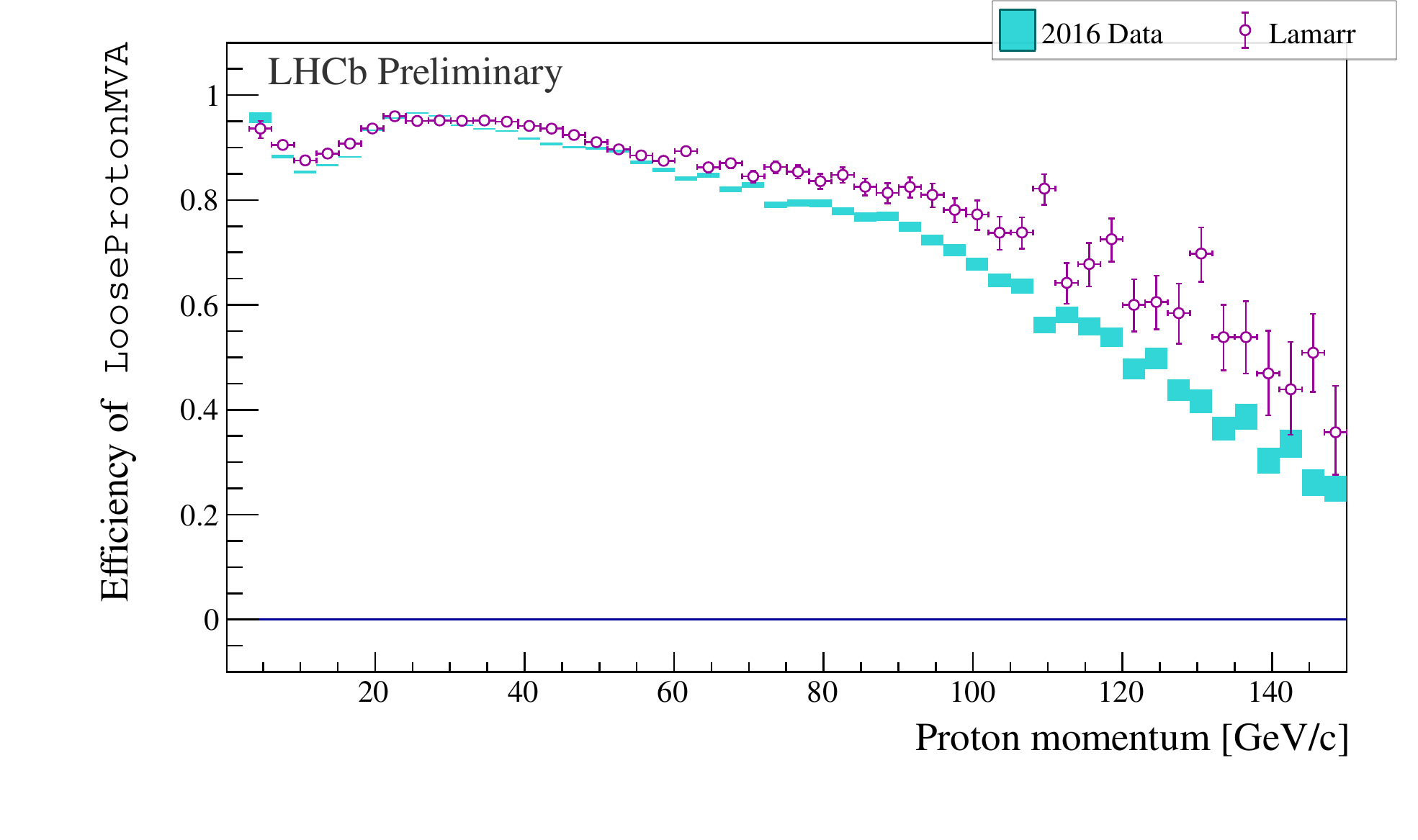} 
    \caption{Loose proton ID }
  \end{subfigure}
   \begin{subfigure}[t]{0.3\textwidth}
    \centering
    \includegraphics[width=1\textwidth]{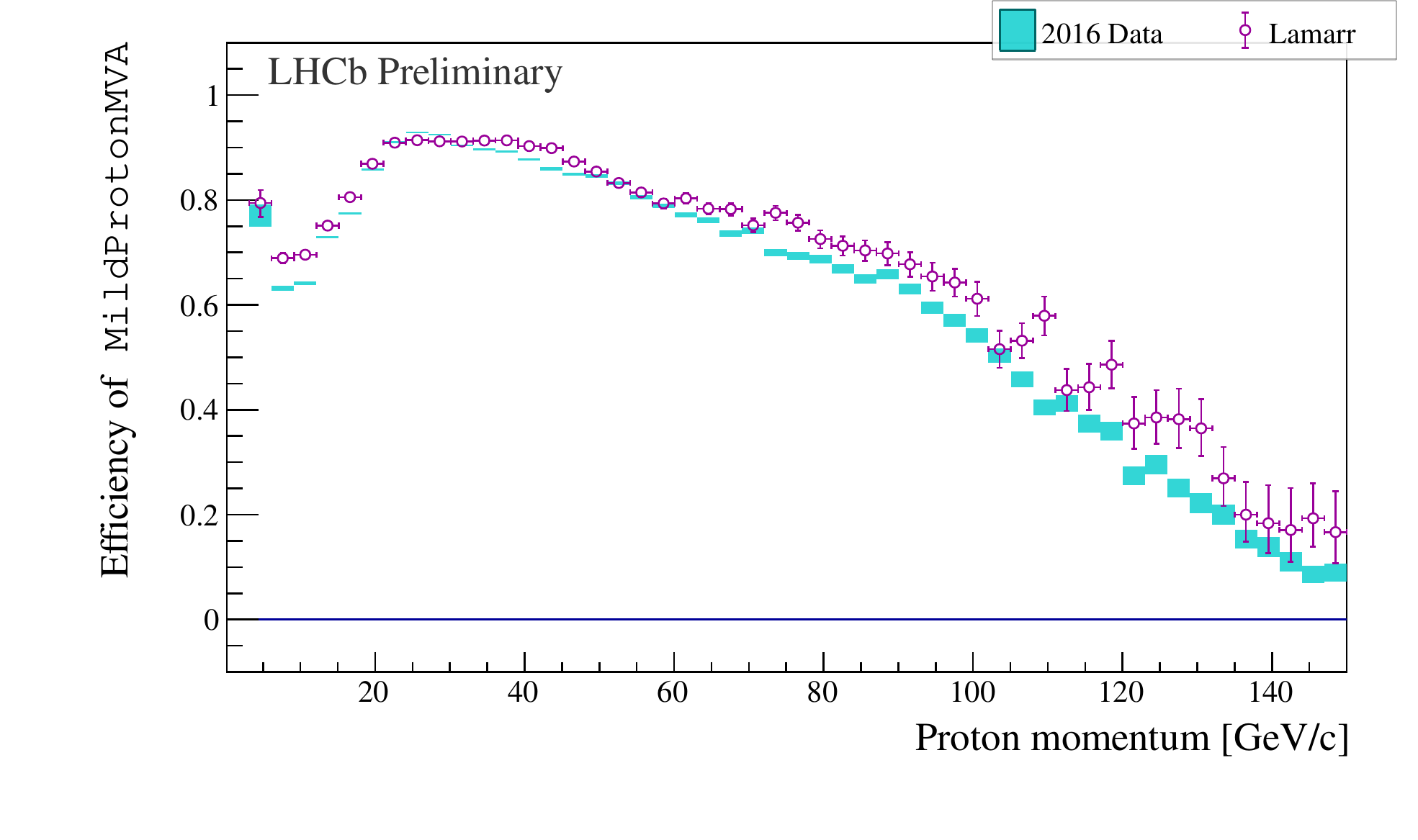}
    \caption{Mild proton ID  }
  \end{subfigure}
  \begin{subfigure}[t]{0.3\textwidth}
    \centering
    \includegraphics[width=1\textwidth]{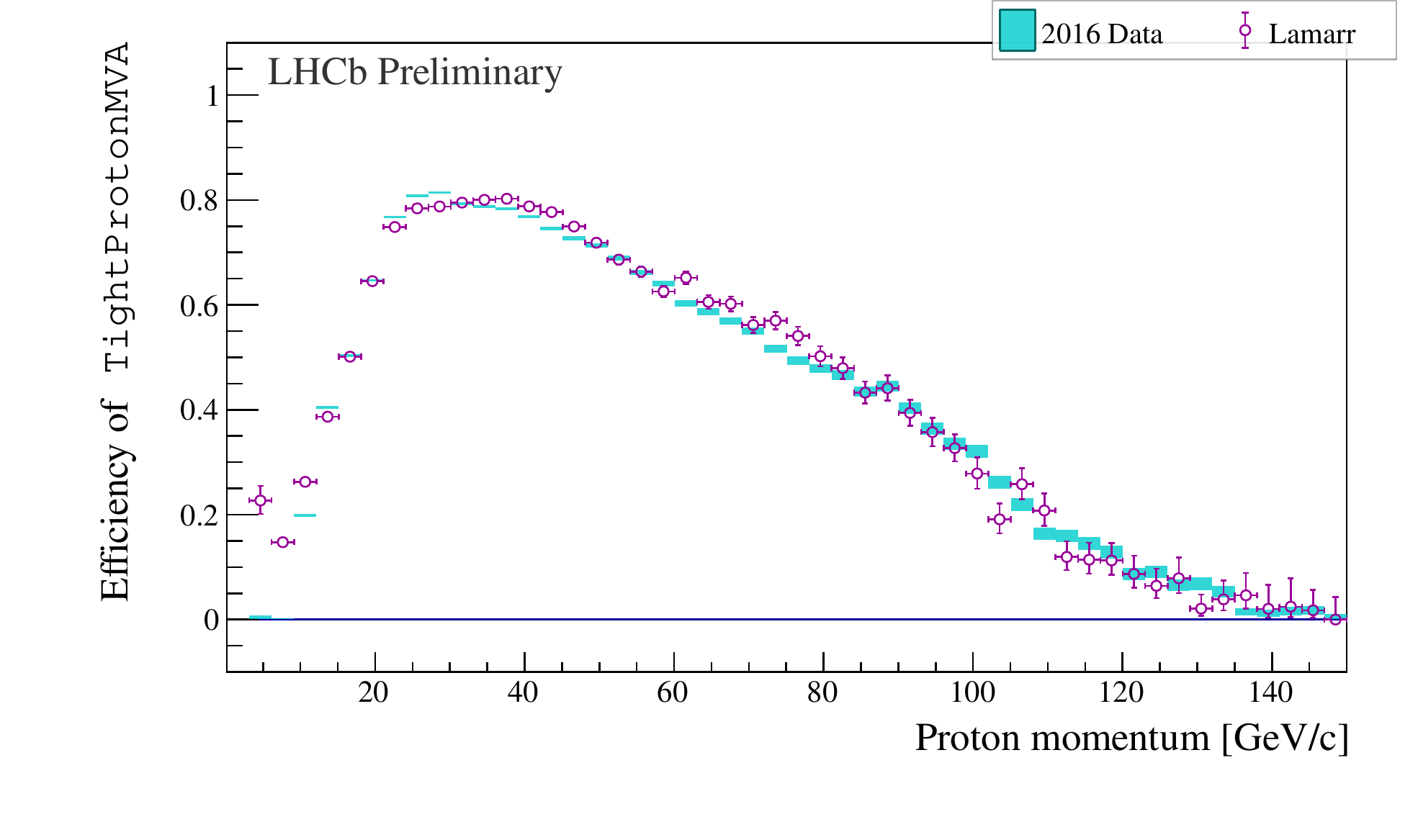} 
    \caption{Tight proton ID }
 \end{subfigure}
  \caption{
Comparing proton identification efficiency. Efficiency of three different requirements on the output of a
Multi-Layer Perceptron trained to identify protons as evaluated on
protons tagged through the decay $\Lambda^0_b \rightarrow \Lambda^+_c
\mu \bar{\nu}_{\mu}$ with $\Lambda^+_c\rightarrow p K^- \pi^+$. The
efficiency is compared, in bins of the proton momentum, for a dataset
selected without introducing bias on the particle identification of
the proton (cyan shaded area), and a Fast Simulation sample where the
rich and the calorimeter responses are modeled through a Generative
Adversarial Network trained using protons from $\Lambda^0\rightarrow p
\pi^-$ decays, only
(purple markers).}
    \label{fig:lemarr}
\end{figure}

A significant increase in total event rate is expected due to upgrades to 
LHC machine and detectors \cite{LHCUpgrade}. The simulation rate will
need to be increased accordingly. However, we can not expect a
significant increase of computational power for computing
hardware. Because computation constraints make it impossible to work harder, we have to work smarter to accommodate the challenge of simulation.   

Using surrogate generative models is one of the possible approaches to
this challenge. It is driven by the observation, that if the physics
detectors has a granularity significantly coarser than the
level of the corresponding \geant simulation, the surrogate model can aggregate micro-level simulation effects into the required macro-level response.

\section{Generative Model for Calorimeter Response Simulation}

The simulation of particle showers in the electromagnetic
calorimeter is the most computationally expensive component of the
Monte Carlo event simulation for the LHCb detector. The relatively
coarse 2D granularity of the calorimeter allows the use of surrogate
generative models to be built on top of the detailed \geant simulation. 
Thus, this approach seems to be promising to speed up the
calorimeter simulation. 
Wasserstein GAN with gradient penalty \cite{wgan} is considered to be a state-of-the-art technique
for image production. We use Wasserstein GAN as a model for generating calorimeter responses. The architecture of this Neural Network and details of training
the generative model are presented in Ref. \cite{chepGAN}.

After the generative model is built and trained,
we compare the original clusters produced by full \geant
simulation with the clusters generated by the trained model for the same
parameters of the incident particles: the same energy, the same
direction, and the same position on the calorimeter 
face. Corresponding images for the four arbitrary parameter sets are
presented in Fig.~\ref{fig:geant_vs_ours}. These images demonstrate very good visual similarity between simulated and generated clusters.

Then, the quantitative evaluation of the proposed
simulation method is performed. While generic evaluation methods for generative
models exist, the evaluation is based on
physics-driven similarity metrics. A few  cluster properties, which essentially drive the cluster properties used in the reconstruction of the calorimeter objects and following physics
analysis, are selected. If the initial particle direction is not perpendicular to the
calorimeter face, the produced cluster is elongated in that direction. 
Therefore, cluster widths in the direction of the
initial particle and in the transverse direction  are considered separately. 
Spatial resolution, which is the distance between the centre of mass of the
cluster and the initial track projection to the shower max depth, is
another important characteristic affecting physics properties of the
cluster.
 Cluster sparsity, which is the fraction of cells with energies above some threshold, reflects marginal low energy properties of the generated clusters. 
These characteristics are presented in Fig.~\ref{fig:ganquality}.

\section{Generative Model for the RICH Particle Identification}

The appropriate generation of the LHCb RICH detector \cite{lhcbTDR} response requires
detailed simulation of the Cherenkov photon production in the body  of
the detector, its transport to the photodetector, including
reflection and refraction on the way, its registering  by the
photodetectors, providing good angular precision for the Cherenkov
photons. Collected signals are used then for testing each charged
particle candidate reconstructed in the tracker against six possible
mass hypotheses: electron, muon, pion, kaon, proton, "below threshold". 
The logarithm of the ratio of the likelihood for each hypothesis, except the pion one, to the likelihood of the pion hypothesis,
\texttt{RichDLL*}, with `\texttt{*}'~standing for  'e', 'mu', 'k', 'p', and 'bt' respectively is associated with every reconstructed charged
track and is used for the particle identification in the following
physics analyses. 

Taking into account symmetry around the beam axis, this chain in fact
converts kinematics of the track, momentum $p$ and pseudorapidity $\eta$ into five
likelihoods for different hypothesis. This gives a possibility to
substitute the directly calculated transfer function which includes
micro-level detector simulation and detector reconstruction for the
effective surrogate model. 
\texttt{RichDLL*} values also depend on the multiplicity in the event, since high track density might lead to a reconstruction algorithm confusion. The proxy variable to a total multiplicity is the number of reconstructed tracks in the event. Thus, the full surrogate
model may have three input parameters: $(p, \eta, N_{tracks})$ where
the latter is a  total number of tracks in the event.

Technical details of this approach are described in
Ref.~\cite{richganACAT}.  The Cramer GAN approach \cite{cramerGAN}
build using fully connected neural network layers presented in Fig.~\ref{fig:richgan}
is used to build and train the surrogate generative model. 
The model was trained using calibration sample of well identified decays in
real data~\cite{calibsamples}. While true id of the particles  is stochastic, thus unknown, the usage of identified decays provides the information of sWeights~\cite{sweights}, which is used in the training process.  

The distributions for \texttt{RichDLLk} values obtained for
pions and kaons directly from the corresponding calibration data samples and generated by the
surrogate model for MC pions and kaons for different regions in $(p,
\eta)$ phase space are presented in Fig.~\ref{fig:RichDLLk}.

  The primary requirement for the surrogate model is to properly reproduce
  a discrimination power of corresponding hypothesis
  estimators \texttt{RichDLL*}. Thus,  Fig.~\ref{fig:DAUCE} represents
  the difference between separation power, ROC AUC, of original and
  surrogate  \texttt{RichDLL*} values, relative to the uncertainty of
  the ROC AUC calculation. This comparison is presented in different bins in momentum
  and pseudorapidity, and demonstrates that the deviation in
  most cases does not exceed 1-2 standard deviations, and is
  essentially unbiased for different bins.

As the results of simulation are used in the subsequent PID algorithm, the ultimate metric ultimate quality metrics for the surrogate  \texttt{RichDLL*}
model is the correct reproduction of identification power. Fig.~\ref{fig:lemarr} presents such a comparison for
the proton identification efficiency for different proton ID
requirements. Demonstrated consistency, especially for the tight
requirements, confirms the feasibility of this approach. 

\section{Conclusions}
In this paper we demonstrated two approaches to significant speedup of
the simulation of two most expansive computationally components of the
LHCb detector: electromagnetic calorimeter and RICH. In the first case the
surrogate generative model is built on top of the highly detailed
\geant response. The surrogate model for the RICH based particle
identification is built on top of real calibration data samples, thus
bypassing the simulation and digitization steps of the MC event
production for this detector completely.

\section{Acknowledgements} 
The research leading to these results has received funding from Russian Science Foundation under grant agreement n$^{\circ}$~19-71-30020.

 \clearpage 
%

\begin{thebibliography}{}
%
%
\bibitem{LHCUpgrade}
LHCb collaboration, CERN-LHCC-2018-007, LHCb-TDR-017.
\bibitem{wgan}
T.C. Wang, M.Y. Liu, J.Y. Zhu, G. Liu, A. Tao, J. Kautz, B. Catanzaro, arXiv preprint
arXiv:1808.06601 (2018).
\bibitem{chepGAN}
V.Chekalina et al., EPJ Web Conf. \textbf{214} 02034 (2019).
\bibitem{lhcbTDR}
Alves Jr A A et al. (LHCb), JINST \textbf{3}, 8005 (2008).
\bibitem{richganACAT}
A Maevskiy et al., Fast Data-Driven Simulation of Cherenkov Detectors
Using Generative Adversarial Networks, 19th International Workshop on
Advanced Computing and Analysis Techniques in Physics Research (in materials),
arXiv preprint arXiv:1905.11825 (2019).
\bibitem{cramerGAN}
M. G. Bellemare et al., arXiv preprint arXiv:1705.10743 (2017).
\bibitem{calibsamples}
R. Aaij et al., EPJ Tech.Instrum. \textbf{6} no.1, 1 (2019)
\bibitem{sweights}
M. Pivk and F. R. Le Diberder, Nucl. Instrum. Meth. \textbf{A555}, 356 (2005)

\end{thebibliography}
%
%

\end{document}